\documentclass[12pt]{article}
\usepackage{amsmath}

\begin{document}

\author{C. Bizdadea\thanks{%
e-mail address: bizdadea@central.ucv.ro}, E. M. Cioroianu\thanks{%
e-mail address: manache@central.ucv.ro}, A. C. Lungu\thanks{%
e-mail address: ac\_lungu@yahoo.com}, S. C. S\u {a}raru\thanks{%
e-mail address: scsararu@central.ucv.ro} \\
Faculty of Physics, University of Craiova\\
13 A. I. Cuza Str., Craiova 200585, Romania}
\title{No cross-interactions between the Weyl graviton and the massless
Rarita-Schwinger field}
\maketitle

\begin{abstract}
The proof of the fact that there are no nontrivial, consistent
cross-couplings that can be added between the Weyl graviton and the massless
Rarita-Schwinger field is accomplished by means of a cohomological approach,
based on the deformation of the solution to the master equation from the
antifield-Becchi-Rouet-Stora-Tyutin (BRST) formalism. The procedure
developed here relies on the assumptions of locality, smoothness,
(background) Lorentz invariance, Poincar\'{e} invariance, and preservation
of the number of derivatives with respect to each field (the last hypothesis
was made only in antighost number zero).

PACS number: 11.10.Ef
\end{abstract}

\section{Introduction}

A key point in the development of local gauge field theory is represented,
without any doubt, by the discovery of the Becchi-Rouet-Stora-Tyutin (BRST)
symmetry~\cite{brs,t} and, in this context, by the cohomological
reformulation of the antifield-BRST symmetry~\cite{coh1,coh2,coh3,coh4,coh5}%
, which allowed the powerful algebraic methods to tackle many issues, like
renormalization and anomalies~\cite{ra1,ra2,ra3,ra4,ra5,ra6,ra7,ra8,ra9,ra10}%
. The applications of the cohomological reformulation of the BRST
approach cover a broad spectrum, including the reformulation of
the problem of consistent interactions among gauge fields as a
cohomological problem of deforming the solution to the master
equation~\cite{def}. The main aim of this paper is to analyze the
construction of consistent interactions that can be introduced
between the linearized limit of Weyl gravity and the massless
Rarita-Schwinger field from the BRST standpoint. Such an analysis
is motivated on the one hand by the remarkable properties of
conformal gravity and supergravity~\cite{fradtseyt}, as well as by
the renewed interest in Weyl gravity~\cite{0007211} in connection
with the ADS/CFT correspondence, and, on the other hand, by some
supergravity models~\cite {SG51,SG52,SG53,SG54}, where the
gravitino represents the supersymmetric partner of the graviton.

Our procedure is based on solving the equations that describe the
deformation of the solution to the master equation by means of specific
cohomological techniques. More precisely, we start from a free model
describing the sum between the linearized limit of Weyl gravity and the
massless Rarita-Schwinger action, and construct its antibracket-antifield
BRST symmetry $s$, which splits as $s=\delta +\gamma $, where $\delta $ is
the Koszul-Tate differential and $\gamma $ represents the exterior
longitudinal derivative. Next, we briefly review the basic equations of the
antibracket-antifield deformation procedure, and then pass to solving the
equation that describes the first-order deformation of the solution to the
master equation. The local form of this equation shows that the
nonintegrated density of the first-order deformation of the solution to the
master equation belongs to the local BRST cohomology in ghost number zero $%
H^{0}(s\vert d) $. The core of the paper is then dedicated to the
computation of $H^{0}(s\vert d) $. Based on this computation, we prove,
under the assumptions of locality, smoothness, (background) Lorentz
invariance, Poincar\'{e} invariance, and preservation of the number of
derivatives with respect to each field (the last hypothesis was made only in
antighost number zero), that there are no nontrivial, consistent
cross-couplings that can be added between the Weyl graviton and the massless
Rarita-Schwinger field. Consequently, the only interactions that can be
added to the Lagrangian action are given by the self-interactions of the
Weyl graviton, studied in detail in~\cite{marcann} from a cohomological
perspective, since the massless Rarita-Schwinger field allows no consistent
self-interactions, as it has been proved in~\cite{boulcqg} also on a
cohomological basis. Our result is in agreement with the fact that it is not
the Rarita-Schwinger field which appears in conformal supergravity, but a
spin $\tfrac{3}{2}$ field, described in the free limit by the action~\cite
{fradtseyt}
\[
S_{0}^{\frac{3}{2}}\left[ \psi _{\mu }\right] \propto \int d^{4}x\left(
\varepsilon ^{\mu \nu \rho \lambda }\bar{\phi}_{\rho }\gamma _{5}\gamma
_{\lambda }\partial _{\mu }\phi _{\nu }\right) ,
\]
where
\[
\phi _{\mu }=\tfrac{1}{3}\gamma ^{\sigma }\left[ \mathrm{i}\left( \partial
_{\sigma }\psi _{\mu }-\partial _{\mu }\psi _{\sigma }\right) +\tfrac{1}{2}%
\gamma _{5}\varepsilon _{\sigma \mu \alpha \beta }\partial ^{\alpha }\psi
^{\beta }\right] ,
\]
so it contains three derivatives instead of only one present in the
Rarita-Schwinger theory.

\section{Free model: Lagrangian formulation and BRST symmetry}

Our starting point is represented by a free Lagrangian action written as the
sum between the linearized Weyl gravity action~\cite{marcann} and the
massless Rarita-Schwinger action~\cite{rs}
\begin{equation}
S_{0}^{L}\left[ h_{\mu \nu },\psi _{\mu }\right] =\tfrac{1}{2}\int
d^{4}x\left( -\varepsilon ^{\mu \nu \rho \lambda }\bar{\psi}_{\mu }\gamma
_{5}\gamma _{\nu }\partial _{\rho }\psi _{\lambda }+\mathcal{W}_{\mu \nu
\alpha \beta }\mathcal{W}^{\mu \nu \alpha \beta }\right) ,  \label{fract}
\end{equation}
where $\mathcal{W}_{\mu \nu \alpha \beta }$ is the linearized Weyl tensor in
four spacetime dimensions, given in terms of the linearized Riemann tensor $%
\mathcal{R}_{\mu \nu \alpha \beta }$ and of its traces by
\begin{equation}
\mathcal{W}_{\mu \nu \alpha \beta }=\mathcal{R}_{\mu \nu \alpha \beta }-%
\tfrac{1}{2}\left( \sigma _{\mu [\alpha }\mathcal{R}_{\beta ]\nu }-\sigma
_{\nu [\alpha }\mathcal{R}_{\beta ]\mu }\right) +\tfrac{1}{6}\mathcal{R}%
\sigma _{\mu [\alpha }\sigma _{\beta ]\nu }.  \label{1}
\end{equation}
Throughout the paper we work with the flat metric of `mostly minus'
signature $\sigma _{\mu \nu }=( +---)$. The notation $[\mu \ldots \nu ]$
signifies full antisymmetry with respect to the indices between brackets
without normalization factors (i.e. the independent terms appear only once
and are not multiplied by overall numerical factors). The linearized Riemann
tensor is expressed by
\begin{eqnarray}
\mathcal{R}_{\mu \nu \alpha \beta } &=&\tfrac{1}{2}\left( \partial _{\mu
}\partial _{\beta }h_{\nu \alpha }+\partial _{\nu }\partial _{\alpha }h_{\mu
\beta }-\partial _{\nu }\partial _{\beta }h_{\mu \alpha }-\partial _{\mu
}\partial _{\alpha }h_{\nu \beta }\right)  \nonumber \\
&\equiv &\tfrac{1}{2}\partial _{[\mu }h_{\nu ][\alpha ,\beta ]},  \label{2}
\end{eqnarray}
while its simple and respectively double traces read as
\begin{equation}
\mathcal{R}_{\mu \nu }=\sigma ^{\alpha \beta }\mathcal{R}_{\mu \alpha \nu
\beta },\qquad \mathcal{R}=\sigma ^{\mu \nu }\mathcal{R}_{\mu \nu }.
\label{3}
\end{equation}
The linearized Weyl tensor can be expressed in terms of the symmetric tensor
$\mathcal{K}_{\mu \nu }$ like
\begin{equation}
\mathcal{W}_{\mu \nu \alpha \beta }=\mathcal{R}_{\mu \nu \alpha \beta
}-\left( \sigma _{\mu [\alpha }\mathcal{K}_{\beta ]\nu }-\sigma _{\nu
[\alpha }\mathcal{K}_{\beta ]\mu }\right) ,  \label{3a}
\end{equation}
where
\begin{equation}
\mathcal{K}_{\mu \nu }=\tfrac{1}{2}\left( \mathcal{R}_{\mu \nu }-\tfrac{1}{6}%
\sigma _{\mu \nu }\mathcal{R}\right) .  \label{3b}
\end{equation}
The spinor-vector $\psi _{\mu }$ has (Majorana) real components and the $%
\gamma $-matrices are in the Majorana representation
\begin{equation}
\gamma _{\mu }^{*}=-\gamma _{\mu },\qquad \gamma _{\mu }^{T}=-\gamma
_{0}\gamma _{\mu }\gamma _{0},\qquad \left( \mu =\overline{0,3}\right) ,
\label{4}
\end{equation}
where $*$ and $T$ in (\ref{4}) signifies the operations of complex
conjugation and respectively of transposition. The theory described by (\ref
{fract}) possesses an irreducible and abelian generating set of gauge
transformations
\begin{equation}
\delta _{\epsilon ,\theta }h_{\mu \nu }=\partial _{(\mu }\epsilon _{\nu
)}+2\sigma _{\mu \nu }\epsilon ,\qquad \delta _{\epsilon ,\theta }\psi _{\mu
}=\partial _{\mu }\theta ,  \label{5and6}
\end{equation}
where the gauge parameters $\epsilon _{\mu }$ and $\epsilon $ are bosonic
and $\theta $ is a fermionic spinor with real components. The scalar gauge
parameter $\epsilon $ is responsible for the so-called conformal invariance
of Weyl theory. The notation $(\mu \nu )$ signifies symmetry with respect to
the indices between parentheses without the factor $1/2$.

In order to construct the BRST symmetry for the model under study we
introduce the ghosts $\eta _{\mu }$, $\xi $ and $C$ respectively associated
with the gauge parameters $\epsilon _{\mu }$, $\epsilon $ and $\theta $. The
ghosts $\eta _{\mu }$ and $\xi $ are fermionic, while $C$ is a real, bosonic
spinor. The antifield spectrum is organized into the antifields $\left\{
h^{*\mu \nu },\psi ^{*\mu }\right\} $ of the original fields $\left\{ h_{\mu
\nu },\psi _{\mu }\right\} $ together with those of the ghosts $\left\{ \eta
^{*\mu },\xi ^{*},C^{*}\right\} $, of statistics opposite to that of the
associated fields/ghosts. The antifields $\psi ^{*\mu }$ are spinor-vectors
with real components and $C^{*}$ is a purely imaginary spinor.

Since the gauge generators of the free theory are field independent, it
follows that the BRST differential simply reduces to
\begin{equation}
s=\delta +\gamma ,  \label{7}
\end{equation}
where $\delta $ represents the Koszul-Tate differential, graded by the
antighost number $\text{agh}$ ($\text{agh}(\delta )=-1$), and $\gamma $
stands for the exterior derivative along the gauge orbits, whose degree is
named pure ghost number $\text{pgh}$ ($\text{pgh}(\gamma )=1$). These two
degrees do not interfere ($\text{pgh}(\delta )=0$, $\text{agh}(\gamma )=0$).
The overall degree that grades the BRST complex is known as the ghost number
$\text{gh} $ and is defined like the difference between the pure ghost
number and the antighost number, such that $\text{gh}(s)=\text{gh}(\delta )=%
\text{gh}(\gamma )=1$. If we denote by
\begin{equation}
\Phi ^{\alpha _{0}}=\left( h_{\mu \nu },\psi _{\mu }\right) ,\qquad \eta
^{\alpha _{1}}=\left( \eta _{\mu },\xi ,C\right)  \label{8}
\end{equation}
the fields and ghosts of the free theory and by
\begin{equation}
\Phi _{\alpha _{0}}^{*}=\left( h^{*\mu \nu },\psi ^{*\mu }\right) ,\qquad
\eta _{\alpha _{1}}^{*}=\left( \eta ^{*\mu },\xi ^{*},C^{*}\right)  \label{9}
\end{equation}
the corresponding antifields, then, according to the standard rules of the
BRST formalism, the corresponding degrees of the generators from the BRST
complex are valued like
\begin{eqnarray}
\text{agh}\left( \Phi ^{\alpha _{0}}\right) &=&0,\qquad \text{agh}\left(
\eta ^{\alpha _{1}}\right) =0,  \label{10} \\
\text{agh}\left( \Phi _{\alpha _{0}}^{*}\right) &=&1,\qquad \text{agh}\left(
\eta _{\alpha _{1}}^{*}\right) =2,  \label{10aaa} \\
\text{pgh}\left( \Phi ^{\alpha _{0}}\right) &=&0,\qquad \text{pgh}\left(
\eta ^{\alpha _{1}}\right) =1,  \label{11aaa} \\
\text{pgh}\left( \Phi _{\alpha _{0}}^{*}\right) &=&0,\qquad \text{pgh}\left(
\eta _{\alpha _{1}}^{*}\right) =0.  \label{11}
\end{eqnarray}
The actions of the differentials $\delta $ and $\gamma $ on the generators (%
\ref{8})--(\ref{9}) from the BRST complex are given by
\begin{equation}
\delta h^{*\mu \nu }=2\partial _{\alpha }\partial _{\beta }\mathcal{W}^{\mu
\alpha \nu \beta },\qquad \delta \psi ^{*\mu }=-\varepsilon ^{\mu \nu \rho
\lambda }\partial _{\nu }\bar{\psi}_{\rho }\gamma _{5}\gamma _{\lambda },
\label{12}
\end{equation}
\begin{equation}
\delta \eta ^{*\mu }=-2\partial _{\nu }h^{*\mu \nu },\qquad \delta \xi
^{*}=2h^{*},\qquad \delta C^{*}=\partial _{\mu }\psi ^{*\mu },  \label{13}
\end{equation}
\begin{equation}
\delta \Phi ^{\alpha _{0}}=0,\qquad \delta \eta ^{\alpha _{1}}=0,  \label{14}
\end{equation}
\begin{equation}
\gamma \Phi _{\alpha _{0}}^{*}=0,\qquad \gamma \eta _{\alpha _{1}}^{*}=0,
\label{15}
\end{equation}
\begin{equation}
\gamma h_{\mu \nu }=\partial _{(\mu }\eta _{\nu )}+2\sigma _{\mu \nu }\xi
,\qquad \gamma \psi _{\mu }=\partial _{\mu }C,  \label{16}
\end{equation}
\begin{equation}
\gamma \eta _{\mu }=\gamma \xi =0,\qquad \gamma C=0.  \label{17}
\end{equation}
The notation $h^{*}$ signifies the trace of $h^{*\mu \nu }$, $h^{*}=\sigma
_{\mu \nu }h^{*\mu \nu }$. The BRST differential is known to have a
canonical action in a structure named antibracket and denoted by the symbol $%
( ,)$ ($s\cdot =(\cdot ,\bar{S})$), which is obtained by decreeing the
fields and ghosts respectively conjugated to the corresponding antifields.
The generator of the BRST symmetry is a bosonic functional, of ghost number
zero, which is solution to the classical master equation $\left( \bar{S},%
\bar{S}\right) =0$. In our case the solution to the master equation reads as
\begin{equation}
\bar{S}=S_{0}^{L}\left[ h_{\mu \nu },\psi _{\mu }\right] +\int d^{4}x\left[
h^{*\mu \nu }\left( \partial _{(\mu }\eta _{\nu )}+2\sigma _{\mu \nu }\xi
\right) +\psi ^{*\mu }\partial _{\mu }C\right] .  \label{18}
\end{equation}

\section{Deformation of the master equation: a brief review}

We begin with a ``free'' gauge theory, described by a Lagrangian action $%
S_{0}^{L}\left[ \Phi ^{\alpha _{0}}\right] $, invariant under some gauge
transformations
\begin{equation}
\delta _{\epsilon }\Phi ^{\alpha _{0}}=Z_{\;\;\alpha _{1}}^{\alpha
_{0}}\epsilon ^{\alpha _{1}},\qquad \frac{\delta S_{0}^{L}}{\delta \Phi
^{\alpha _{0}}}Z_{\;\;\alpha _{1}}^{\alpha _{0}}=0,  \label{2.1}
\end{equation}
and consider the problem of constructing consistent interactions among the
fields $\Phi ^{\alpha _{0}}$ such that the couplings preserve the field
spectrum and the original number of gauge symmetries. This matter is
addressed by means of reformulating the problem of constructing consistent
interactions as a deformation problem of the solution to the master equation
corresponding to the ``free'' theory \cite{def}. Such a reformulation is
possible due to the fact that the solution to the master equation contains
all the information on the gauge structure of the theory. If an interacting
gauge theory can be consistently constructed, then the solution $\bar{S}$ to
the master equation associated with the ``free'' theory, $\left( \bar{S},%
\bar{S}\right) =0$, can be deformed into a solution $S$
\begin{eqnarray}
\bar{S}\rightarrow S &=&\bar{S}+gS_{1}+g^{2}S_{2}+\cdots  \nonumber \\
&=&\bar{S}+g\int d^{D}x\,a+g^{2}\int d^{D}x\,b+\cdots  \label{2.2}
\end{eqnarray}
of the master equation for the deformed theory
\begin{equation}
\left( S,S\right) =0,  \label{2.3}
\end{equation}
such that both the ghost and antifield spectra of the initial theory are
preserved. The equation (\ref{2.3}) splits, according to the various orders
in the coupling constant (deformation parameter) $g$, into
\begin{eqnarray}
\left( \bar{S},\bar{S}\right) &=&0,  \label{2.4} \\
2\left( S_{1},\bar{S}\right) &=&0,  \label{2.5} \\
2\left( S_{2},\bar{S}\right) +\left( S_{1},S_{1}\right) &=&0,  \label{2.6} \\
\left( S_{3},\bar{S}\right) +\left( S_{1},S_{2}\right) &=&0,  \label{2.7} \\
&&\vdots  \nonumber
\end{eqnarray}

The equation (\ref{2.4}) is fulfilled by hypothesis. The next one requires
that the first-order deformation of the solution to the master equation, $%
S_{1}$, is a co-cycle of the ``free'' BRST differential $s\cdot =(\cdot ,%
\bar{S})$. However, only cohomologically nontrivial solutions to (\ref{2.5})
should be taken into account, as the BRST-exact ones can be eliminated by
some (in general nonlinear) field redefinitions. This means that $S_{1}$
pertains to the ghost number zero cohomological space of $s$, $H^{0}(s)$,
which is generically nonempty due to its isomorphism to the space of
physical observables of the ``free'' theory. It has been shown (on behalf of
the triviality of the antibracket map in the cohomology of the BRST
differential) that there are no obstructions in finding solutions to the
remaining equations ((\ref{2.6})--(\ref{2.7}), etc.). However, the resulting
interactions may be nonlocal, and there might even appear obstructions if
one insists on their locality. The analysis of these obstructions can be
done with the help of cohomological techniques.

\section{Consistent interactions between the Weyl graviton and the massless
Rarita-Schwinger field}

\subsection{Standard material: $H(\gamma )$ and $H(\delta \vert d)$}

The aim of this paper is the investigation of the effective couplings that
can be introduced between the Weyl graviton and the massless
Rarita-Schwinger field. This matter is addressed in the context of the
antifield-BRST deformation procedure described in the above and relies on
computing the solutions to the equations (\ref{2.5})--(\ref{2.7}), etc.,
with the help of the BRST cohomology. For obvious reasons, we consider only
smooth, local, (background) Lorentz invariant and, moreover, Poincar\'{e}
invariant quantities (i.e. we do not allow explicit dependence on the
spacetime coordinates). If we make the notation $S_{1}=\int d^{4}x\,a$, with
$a$ a local function, then the equation (\ref{2.5}), which we have seen that
controls the first-order deformation, takes the local form
\begin{equation}
sa=\partial _{\mu }m^{\mu },\qquad \text{gh}(a)=0,\qquad \varepsilon (a)=0,
\label{3.1}
\end{equation}
for some local $m^{\mu }$ and it shows that the nonintegrated density of the
first-order deformation pertains to the local cohomology of $s$ in ghost
number zero, $a\in H^{0}(s\vert d)$, where $d$ denotes the exterior
spacetime differential. The solution to the equation (\ref{3.1}) is unique
up to $s$-exact pieces plus divergences
\begin{eqnarray}
a &\rightarrow &a+sb+\partial _{\mu }n^{\mu },  \label{3.1a} \\
\text{gh}(b) &=&-1,\qquad \varepsilon (b)=1,\qquad \text{gh}\left( n^{\mu
}\right) =0,\qquad \varepsilon \left( n^{\mu }\right) =0.  \nonumber
\end{eqnarray}
At the same time, if the general solution of (\ref{3.1}) is found to be
completely trivial, $a=sb+\partial _{\mu }n^{\mu }$, then it can be made to
vanish $a=0$.

In order to analyze the equation (\ref{3.1}), we develop $a$ according to
the antighost number
\begin{equation}
a=\sum\limits_{i=0}^{I}a_{i},\qquad \text{agh}(a_{i})=i,\qquad \text{gh}%
(a_{i})=0,\qquad \varepsilon (a_{i})=0,  \label{3.2}
\end{equation}
and assume, without loss of generality, that the decomposition (\ref{3.2})
stops at some finite value of $I$. This can be shown, for instance, like in~%
\cite{gen2} (Section 3), under the sole assumption that the interacting
Lagrangian at the first order in the coupling constant, $a_{0}$, has a
finite, but otherwise arbitrary derivative order. Replacing the
decomposition (\ref{3.2}) into the equation (\ref{3.1}) and projecting it on
various values of the antighost number, we obtain the tower of equations
\begin{eqnarray}
\gamma a_{I} &=&\partial _{\mu }\stackrel{(I)}{m}^{\mu },  \label{3.3} \\
\delta a_{I}+\gamma a_{I-1} &=&\partial _{\mu }\stackrel{(I-1)}{m}^{\mu },
\label{3.4} \\
\delta a_{i}+\gamma a_{i-1} &=&\partial _{\mu }\stackrel{(i-1)}{m}^{\mu
},\qquad (1\leq i\leq I-1),  \label{3.5}
\end{eqnarray}
where $\left( \stackrel{(i)}{m}^{\mu }\right) _{i=\overline{0,I}}$ are some
local currents with $\text{agh}\left( \stackrel{(i)}{m}^{\mu }\right) =i$.
It can be proved that the equation (\ref{3.3}) can be replaced in strictly
positive antighost numbers by
\begin{equation}
\gamma a_{I}=0,\qquad I>0.  \label{3.6}
\end{equation}
The proof of this result is standard material and can be found for instance
in~\cite{marcann,boulcqg,gen2,multi,lingr,hepth04}. Due to the second-order
nilpotency of $\gamma $ ($\gamma ^{2}=0$), the solution to the equation (\ref
{3.6}) is clearly unique up to $\gamma $-exact contributions
\begin{eqnarray}
&& a_{I}\rightarrow a_{I}+\gamma b_{I},  \label{r68} \\
&& \text{agh}(b_{I})=I,\qquad \text{pgh}(b_{I})=I-1,\qquad \varepsilon
(b_{I})=1.  \nonumber
\end{eqnarray}
Meanwhile, if it turns out that $a_{I}$ reduces to $\gamma $-exact terms
only, $a_{I}=\gamma b_{I}$, then it can be made to vanish, $a_{I}=0$. In
other words, the nontriviality of the first-order deformation $a$ is
translated at its highest antighost number component into the requirement
that $a_{I}\in H^{I}( \gamma )$, where $H^{I}(\gamma ) $ denotes the
cohomology of the exterior longitudinal derivative $\gamma $ in pure ghost
number equal to $I$. So, in order to solve the equation (\ref{3.1})
(equivalent with (\ref{3.6}) and (\ref{3.4})--(\ref{3.5})), we need to
compute the cohomology of $\gamma $, $H(\gamma )$, and, as it will be made
clear below, also the local cohomology of $\delta $, $H(\delta \vert d)$.

In order to determine the cohomology $H(\gamma )$, we split the differential
$\gamma $ into two pieces
\begin{equation}
\gamma =\gamma _{W}+\gamma _{\text{RS}},  \label{3.7}
\end{equation}
where $\gamma _{W}$ acts nontrivially only on the fields/ghosts from the
Weyl sector, while $\gamma _{\text{RS}}$ does the same thing, but with
respect to the Rarita-Schwinger sector. From the above splitting it follows
that the nilpotency of $\gamma $ is equivalent to the nilpotency and
anticommutation of its components
\begin{equation}
\left( \gamma _{W}\right) ^{2}=0=\left( \gamma _{\text{RS}}\right)
^{2},\qquad \gamma _{W}\gamma _{\text{RS}}+\gamma _{\text{RS}}\gamma _{W}=0,
\label{3.8}
\end{equation}
so finally we find the isomorphism
\begin{equation}
H(\gamma )=H(\gamma _{W})\otimes H(\gamma _{\text{RS}}).  \label{3.9}
\end{equation}
Using the results from the literature concerning the cohomologies $H(\gamma
_{W})$ and $H(\gamma _{\text{RS}})$~\cite{marcann,boulcqg} we can state that
$H(\gamma )$ is generated on the one hand by $\Phi _{\alpha _{0}}^{*}$, $%
\eta _{\alpha _{1}}^{*}$, $\partial _{[\mu }\psi _{\nu ]}$ and $\mathcal{W}%
_{\mu \nu \alpha \beta }$ as well as by their spacetime derivatives and, on
the other hand, by the ghosts $C$, $\eta _{\mu }$, $\partial _{[\mu }\eta
_{\nu ]}$, $\xi $ and $\partial _{\mu }\xi $. So, the most general (and
nontrivial), local solution to (\ref{3.6}) can be written, up to $\gamma $%
-exact contributions, as
\begin{equation}
a_{I}=\alpha _{I}\left( \left[ \partial _{[\mu }\psi _{\nu ]}\right] ,\left[
\mathcal{W}_{\mu \nu \alpha \beta }\right] ,\left[ \Phi _{\alpha
_{0}}^{*}\right] ,\left[ \eta _{\alpha _{1}}^{*}\right] \right) \omega
^{I}\left( C,\eta _{\mu },\partial _{[\mu }\eta _{\nu ]},\xi ,\partial _{\mu
}\xi \right) ,  \label{3.10}
\end{equation}
where the notation $f([q])$ means that $f$ depends on $q$ and its
derivatives up to a finite order, while $\omega ^{I}$ denotes the elements
of a basis in the space of polynomials with pure ghost number $I $ in the
corresponding ghosts and some of their first-order derivatives. The objects $%
\alpha _{I}$ (obviously nontrivial in $H^{0}(\gamma )$) were taken to have a
bounded number of derivatives, and therefore they are polynomials in the
antifields $\Phi _{\alpha _{0}}^{*}$ and $\eta _{\alpha _{1}}^{*}$, in the
linearized Weyl tensor $\mathcal{W}_{\mu \nu \alpha \beta }$, in the objects
$\partial _{[\mu }\psi _{\nu ]}$, as well as in their derivatives. They are
required to fulfill the property $\text{agh}(\alpha _{I})=I$ in order to
ensure that the ghost number of $a_{I}$ is equal to zero. Due to their $%
\gamma $-closeness, $\gamma \alpha _{I}=0$, $\alpha _{I}$ will be
called ``invariant polynomials''. In zero antighost number, the
invariant polynomials are polynomials in the linearized Weyl
tensor $\mathcal{W}_{\mu \nu \alpha \beta }$, in the objects
$\partial _{[\mu }\psi _{\nu ]}$, and also in their derivatives.

Substituting (\ref{3.10}) in (\ref{3.4}) we obtain that a necessary (but not
sufficient) condition for the existence of (nontrivial) solutions $a_{I-1}$
is that the invariant polynomials $\alpha _{I}$ are (nontrivial) objects
from the local cohomology of Koszul-Tate differential $H(\delta \vert d)$ in
antighost number $I>0$ and pure ghost number zero,
\begin{eqnarray}
&& \delta \alpha _{I}=\partial _{\mu }\stackrel{(I-1)}{j}^{\mu },
\label{3.10a} \\
&& \text{agh}\left( \stackrel{(I-1)}{j}^{\mu }\right) =I-1,\qquad \text{pgh}%
\left( \stackrel{(I-1)}{j}^{\mu }\right) =0.  \nonumber
\end{eqnarray}
We recall that the local cohomology $H(\delta \vert d)$ is completely
trivial in both strictly positive antighost \textit{and} pure ghost numbers
(for instance, see~\cite{gen1,gen11}, Theorem 5.4 and~\cite{commun1}). Using
the fact that the Cauchy order of the free theory under study is equal to
two together with the general results from~\cite{gen1,gen11}, according to
which the local cohomology of the Koszul-Tate differential in pure ghost
number zero is trivial in antighost numbers strictly greater than its Cauchy
order, we can state that
\begin{equation}
H_{J}(\delta \vert d)=0\qquad \text{for all}\qquad J>2,  \label{3.11}
\end{equation}
where $H_{J}(\delta \vert d)$ denotes the local cohomology of the
Koszul-Tate differential in antighost number $J$ and in zero pure ghost
number. It is quite reasonable to assume that if the invariant polynomial $%
\alpha _{J}$, with $\text{agh}(\alpha _{J})=J\geq 2$, is trivial in $%
H_{J}(\delta \vert d)$, then it can be taken to be trivial also in $H_{J}^{%
\text{inv}}(\delta \vert d)$%
\begin{equation}
\left( \alpha _{J}=\delta b_{J+1}+\partial _{\mu }\stackrel{(J)}{c}^{\mu
},\qquad \text{agh}\left( \alpha _{J}\right) =J\geq 2\right) \Rightarrow
\left( \alpha _{J}=\delta \beta _{J+1}+\partial _{\mu }\stackrel{(J)}{\gamma
}^{\mu }\right) ,  \label{3.12ax}
\end{equation}
with both $\beta _{J+1}$ and $\stackrel{(J)}{\gamma }^{\mu }$ invariant
polynomials. Here, $H_{J}^{\text{inv}}(\delta \vert d)$ denotes the
invariant characteristic cohomology in antighost number $J$ (the local
cohomology of the Koszul-Tate differential in the space of invariant
polynomials). This assumption is based on what happens in many gauge
theories~\cite{marcann,boulcqg,gen2,multi,lingr,hepth04,noijhep}. The
results (\ref{3.11})--(\ref{3.12ax}) yield the conclusion that
\begin{equation}
H_{J}^{\text{inv}}(\delta \vert d)=0,\qquad \text{for all}\qquad J>2.
\label{3.12x}
\end{equation}
With the help of the definitions (\ref{12})--(\ref{14}) we observe that both
$H_{2}(\delta \vert d)$ in pure ghost number zero and $H_{2}^{\text{inv}%
}(\delta \vert d)$ are spanned only by the undifferentiated antifields
\begin{equation}
H_{2}(\delta \vert d)\text{ and }H_{2}^{\text{inv}}(\delta \vert d):\qquad
\left( C^{*},\eta ^{*\mu }\right) .  \label{3.12b}
\end{equation}
In contrast to the groups $\left( H_{J}(\delta \vert d)\right) _{J\geq 2}$
and $\left( H_{J}^{\text{inv}}(\delta \vert d)\right) _{J\geq 2}$, which are
finite-dimensional, the cohomology $H_{1}(\delta \vert d)$ in pure ghost
number zero, that is related to global symmetries and ordinary conservation
laws, is infinite-dimensional since the theory is free. Fortunately, it will
not be needed in the sequel.

The previous results on $H(\delta |d)$ and $H^{\text{inv}}(\delta |d)$ in
strictly positive antighost numbers are important because they control the
obstructions to removing the antifields from the first-order deformation.
This statement is also standard material and can be done like in~\cite
{marcann,boulcqg,gen2,multi,lingr,hepth04,noijhep}. Its proof is mainly
based on the formulas (\ref{3.11})--(\ref{3.12x}) and relies on the fact
that we can successively eliminate all the pieces of antighost number
strictly greater than two from the nonintegrated density of the first-order
deformation by adding only trivial terms, so we can take, without loss of
nontrivial objects, the condition $I\leq 2$ in the decomposition (\ref{3.2}%
). In addition, the last representative is of the form (\ref{3.10}), where
the invariant polynomial is necessarily a nontrivial object from $H_{2}^{%
\text{inv}}(\delta |d)$ for $I=2$, respectively from $H_{1}(\delta |d)$ for $%
I=1$.

\subsection{Case $I=2$\label{i2}}

In the case $I=2$ the nonintegrated density of the first-order deformation (%
\ref{3.2}) becomes
\begin{equation}
a=a_{0}+a_{1}+a_{2}.  \label{3.12}
\end{equation}
We can further decompose $a$ in a natural manner as
\begin{equation}
a=a^{(W)}+a^{(\text{RS})}+a^{(\text{int})},  \label{3.12a}
\end{equation}
where $a^{(W)}$ contains only fields/ghosts/antifields from the Weyl sector,
$a^{(\text{RS})}$ is strictly related to the Rarita-Schwinger theory and $%
a^{(\text{int})}$ describes the cross-interactions between the two theories,
so it effectively mixes both sectors. Each of the three components satisfies
an individual equation of the type (\ref{3.1}) and admits a decomposition
similar to (\ref{3.12})
\begin{equation}
a_{k}=a_{k}^{(W)}+a_{k}^{(\text{RS})}+a_{k}^{(\text{int})},\qquad (k=0,1,2),
\label{3.13}
\end{equation}
so each type of deformation is subject to a set of equations of the form (%
\ref{3.6}) and (\ref{3.4})--(\ref{3.5}) for $I=2$. We are interested only in
the term $a_{k}^{(\text{int})}$ since the others merely describe the
self-interactions of Weyl and respectively of massless Rarita-Schwinger
theory, which have already been studied in the literature. The
self-interactions of Weyl theory are known to describe the Weyl gravity
action~\cite{marcann}, while the Rarita-Schwinger model leads to no
consistent self-interactions~\cite{boulcqg}. Using the formula (\ref{3.10})
for pure ghost number two and the result (\ref{3.12b}), we deduce that the
most general, nontrivial element $a_{2}^{(\text{int})}$ (as solution to the
equation $\gamma a_{2}^{(\text{int})}=0$) is
\begin{eqnarray}
a_{2}^{(\text{int})} &=&k_{1}\eta ^{*\mu }\bar{C}\gamma _{\mu
}C+k_{2}\partial _{[\mu }\eta _{\nu ]}C^{*}\gamma ^{[\mu }\gamma ^{\nu
]}C+k_{3}\eta _{\mu }C^{*}\gamma ^{\mu }C  \nonumber \\
&&+k_{4}\left( \partial _{\mu }\xi \right) C^{*}\gamma ^{\mu }C+k_{5}\xi
C^{*}C,  \label{3.14}
\end{eqnarray}
with $\left( k_{a}\right) _{a=\overline{1,5}}$ some arbitrary complex
constants ($k_{2}$ and $k_{5}$ are real numbers and the others are purely
imaginary). On behalf of the definitions (\ref{12})--(\ref{17}), from (\ref
{3.14}) we get
\begin{eqnarray}
\delta a_{2}^{(\text{int})} &=&\partial _{\mu }j_{1}^{\mu }+\gamma b_{1}
\nonumber \\
&&-\tfrac{1}{2}k_{3}\partial _{[\mu }\eta _{\nu ]}\psi ^{*\mu }\gamma ^{\nu
}C+k_{3}\xi \psi ^{*\mu }\gamma _{\mu }C  \nonumber \\
&&-\partial _{\nu }\xi \left( 4k_{2}\psi _{\mu }^{*}\gamma ^{[\mu }\gamma
^{\nu ]}C+k_{5}\psi ^{*\nu }C\right) ,  \label{3.15}
\end{eqnarray}
where
\begin{eqnarray}
j_{1}^{\mu } &=&-2k_{1}h^{*\mu \nu }\bar{C}\gamma _{\nu }C+k_{2}\partial
_{[\alpha }\eta _{\beta ]}\psi ^{*\mu }\gamma ^{[\alpha }\gamma ^{\beta ]}C
\nonumber \\
&&+\left( k_{4}\left( \partial _{\alpha }\xi \right) +k_{3}\eta _{\alpha
}\right) \psi ^{*\mu }\gamma ^{\alpha }C+k_{5}\xi \psi ^{*\mu }C,
\label{3.16}
\end{eqnarray}
\begin{eqnarray}
b_{1} &=&4k_{1}h^{*\mu \nu }\bar{C}\gamma _{\mu }\psi _{\nu }-k_{2}\partial
_{[\mu }\eta _{\nu ]}\psi _{\rho }^{*}\gamma ^{[\mu }\gamma ^{\nu ]}\psi
^{\rho }  \nonumber \\
&&-\left[ k_{3}\eta _{\mu }+k_{4}\left( \partial _{\mu }\xi \right) \right]
\psi _{\rho }^{*}\gamma ^{\mu }\psi ^{\rho }  \nonumber \\
&&-k_{5}\xi \psi _{\rho }^{*}\psi ^{\rho }+\left( k_{4}\mathcal{K}_{\mu \nu
}-\tfrac{1}{2}k_{3}h_{\mu \nu }\right) \psi ^{*\mu }\gamma ^{\nu }C
\nonumber \\
&&-k_{2}\partial _{[\mu }h_{\nu ]\rho }\psi ^{*\rho }\gamma ^{[\mu }\gamma
^{\nu ]}C.  \label{3.17}
\end{eqnarray}
By means of the relation (\ref{3.15}), we observe that the existence of $%
a_{1}^{(\text{int})}$ as solution to the equation $\delta a_{2}^{(\text{int}%
)}+\gamma a_{1}^{(\text{int})}=\partial _{\mu }\stackrel{(1)}{m}^{(\text{int}%
)\mu }$ requires that
\begin{equation}
k_{2}=k_{3}=k_{5}=0.  \label{3.18}
\end{equation}
Inserting (\ref{3.18}) in (\ref{3.14}) and (\ref{3.15}), we have that the
pieces of antighost number two and one from $a^{(\text{int})}$ respectively
read as
\begin{equation}
a_{2}^{(\text{int})}=k_{1}\eta ^{*\mu }\bar{C}\gamma _{\mu }C+k_{4}\left(
\partial _{\mu }\xi \right) C^{*}\gamma ^{\mu }C,  \label{3.19}
\end{equation}
\begin{equation}
a_{1}^{(\text{int})}=-4k_{1}h^{*\mu \nu }\bar{C}\gamma _{\mu }\psi _{\nu
}+k_{4}\left[ \left( \partial _{\mu }\xi \right) \psi _{\rho }^{*}\gamma
^{\mu }\psi ^{\rho }-\mathcal{K}_{\mu \nu }\psi ^{*\mu }\gamma ^{\nu
}C\right] .  \label{3.20}
\end{equation}
After some computation we find that
\begin{equation}
\delta a_{1}^{(\text{int})}=\partial _{\mu }j_{0}^{\mu }+\gamma b_{0}+c_{0},
\label{3.21}
\end{equation}
where
\begin{eqnarray}
j_{0}^{\mu } &=&-8k_{1}\left( \partial _{\beta }\mathcal{W}^{\mu \alpha \nu
\beta }\right) \bar{\psi}_{\alpha }\gamma _{\nu }C+k_{4}\varepsilon ^{\mu
\nu \rho \lambda }\left[ \tfrac{1}{2}\left( \partial _{\alpha }\xi \right)
\bar{\psi}_{\nu }\gamma _{5}\gamma _{\rho }\gamma ^{\alpha }\psi _{\lambda
}\right.   \nonumber \\
&&\left. +\xi \left( \partial _{\nu }\bar{\psi}_{\rho }\right) \gamma
_{5}\psi _{\lambda }-\mathcal{K}_{\nu \alpha }\bar{\psi}_{\rho }\gamma
_{5}\gamma _{\lambda }\gamma ^{\alpha }C\right] ,  \label{3.22}
\end{eqnarray}
\begin{equation}
b_{0}=-4k_{1}\left( \partial _{\beta }\mathcal{W}^{\mu \alpha \nu \beta
}\right) \bar{\psi}_{\mu }\gamma _{\nu }\psi _{\alpha }-\tfrac{1}{2}k_{4}%
\mathcal{K}_{\mu \alpha }\varepsilon ^{\mu \nu \rho \lambda }\bar{\psi}_{\nu
}\gamma _{5}\gamma _{\rho }\gamma ^{\alpha }\psi _{\lambda },  \label{3.23}
\end{equation}
\begin{eqnarray}
c_{0} &=&4k_{1}\left( \partial _{\beta }\mathcal{W}^{\mu \alpha \nu \beta
}\right) \partial _{[\mu }\bar{\psi}_{\alpha ]}\gamma _{\nu }C  \nonumber \\
&&+k_{4}\varepsilon ^{\mu \nu \rho \lambda }\left[ \left( \partial _{\mu }%
\mathcal{K}_{\nu \alpha }\right) \bar{\psi}_{\rho }\gamma _{5}\gamma
_{\lambda }\gamma ^{\alpha }C+\xi \left( \partial _{\mu }\bar{\psi}_{\nu
}\right) \gamma _{5}\left( \partial _{\rho }\psi _{\lambda }\right) \right] .
\label{3.24}
\end{eqnarray}
Combining (\ref{3.21}) with (\ref{3.24}), we observe that the existence of $%
a_{0}^{(\text{int})}$ as solution to the equation $\delta a_{1}^{(\text{int}%
)}+\gamma a_{0}^{(\text{int})}=\partial _{\mu }\stackrel{(0)}{m}^{(\text{int}%
)\mu }$ implies that $c_{0}=0$, which is equivalent to the vanishing of the
remaining constants
\begin{equation}
k_{1}=k_{4}=0.  \label{3.24a}
\end{equation}
Replacing (\ref{3.24a}) in (\ref{3.19})--(\ref{3.20}), we conclude there is
no nontrivial first-order cross-coupling that stops at antighost number two.
It is remarkable to note that the impossibility of consistent, nontrivial,
first-order deformations that end at antighost number two was obtained \emph{%
without} imposing any restriction on the derivative order of the interacting
Lagrangian.

\subsection{Case $I=1$\label{i1}}

The next step is to investigate whether there exist nontrivial,
cross-coupling first-order deformations that end at antighost number one ($%
I=1$)
\begin{equation}
a^{(\text{int})}=a_{0}^{(\text{int})}+a_{1}^{(\text{int})},  \label{3.25}
\end{equation}
with $a_{1}^{(\text{int})}$ from $H^{1}(\gamma )$. According to (\ref{3.10}%
), we deduce that the most general form of $a_{1}^{(\text{int})}$ that might
provide effective cross-interactions is written like
\begin{equation}
a_{1}^{(\text{int})}=\psi ^{*\mu }\left( \tilde{M}_{\mu \nu }\partial ^{\nu
}\xi +\bar{M}_{\mu }^{\;\;\alpha \beta }\partial _{[\alpha }\eta _{\beta ]}+%
\hat{M}_{\mu }\xi +\hat{M}_{\mu \nu }\eta ^{\nu }\right) +h^{*\mu \nu }\bar{C%
}M_{\mu \nu },  \label{3.26}
\end{equation}
where $\tilde{M}_{\mu \nu }$, $\bar{M}_{\mu }^{\;\;\alpha \beta }$, $\hat{M}%
_{\mu }$, $\hat{M}_{\mu \nu }$ and $M_{\mu \nu }$ are spinor-tensors
depending on the original fields, which are gauge-invariant, fermionic
objects (each of them can be generically written like $M_{\Delta }=U_{\Delta
}^{\alpha \beta }\partial _{[\alpha }\psi _{\beta ]}$, where $U_{\Delta
}^{\alpha \beta }$ are bosonic matrices, possibly involving $\mathcal{W}%
^{\mu \alpha \nu \beta }$). All the matrices appearing in (\ref{3.26}) must
be purely imaginary. Direct calculations based on the definitions (\ref{12}%
)--(\ref{17}) lead to
\begin{equation}
\delta a_{1}^{(\text{int})}=\partial _{\mu }m_{0}^{\mu }+\gamma e_{0}+f_{0},
\label{3.27}
\end{equation}
where we made the notations
\begin{eqnarray}
m_{0}^{\mu } &=&\varepsilon ^{\mu \nu \rho \lambda }\bar{\psi}_{\nu }\gamma
_{5}\gamma _{\rho }\left( \tilde{M}_{\lambda \alpha }\partial ^{\alpha }\xi +%
\bar{M}_{\lambda }^{\;\;\alpha \beta }\partial _{[\alpha }\eta _{\beta ]}+%
\hat{M}_{\lambda }\xi +\hat{M}_{\lambda \alpha }\eta ^{\alpha }\right)
\nonumber \\
&&+2\left( \partial _{\beta }\mathcal{W}^{\mu \alpha \nu \beta }\right) \bar{%
C}M_{\alpha \nu }+\mathcal{W}^{\mu \alpha \nu \beta }\bar{C}\partial _{[\nu
}M_{\beta ]\alpha }  \nonumber \\
&&+\varepsilon ^{\mu \nu \rho \lambda }\bar{C}\gamma _{5}\gamma _{\nu
}\left( -\tilde{M}_{\rho \alpha }\mathcal{K}_{\lambda }^{\alpha }+\bar{M}%
_{\rho }^{\;\;\alpha \beta }\partial _{[\alpha }h_{\beta ]\lambda }+\tfrac{1%
}{2}\hat{M}_{\rho \alpha }h_{\lambda }^{\alpha }\right) ,  \label{3.28}
\end{eqnarray}
\begin{eqnarray}
e_{0} &=&\varepsilon ^{\mu \nu \rho \lambda }\bar{\psi}_{\mu }\gamma
_{5}\gamma _{\nu }\left( -\tilde{M}_{\rho \alpha }\mathcal{K}_{\lambda
}^{\alpha }+\bar{M}_{\rho }^{\;\;\alpha \beta }\partial _{[\alpha }h_{\beta
]\lambda }+\tfrac{1}{2}\hat{M}_{\rho \alpha }h_{\lambda }^{\alpha }\right)
\nonumber \\
&&-2\left( \partial _{\beta }\mathcal{W}^{\mu \alpha \nu \beta }\right) \bar{%
\psi}_{\alpha }M_{\mu \nu }-\mathcal{W}^{\mu \alpha \nu \beta }\bar{\psi}%
_{\beta }\partial _{[\mu }M_{\alpha ]\nu },  \label{3.29}
\end{eqnarray}
\begin{eqnarray}
f_{0} &=&-\tfrac{1}{2}\varepsilon ^{\mu \nu \rho \lambda }\bar{\psi}_{\mu
}\gamma _{5}\gamma _{\nu }\left[ \partial _{[\rho }\hat{M}_{\lambda ]\alpha
}\eta ^{\alpha }+\left( \partial _{[\rho }\bar{M}_{\lambda ]}^{\;\;\alpha
\beta }+\tfrac{1}{4}\hat{M}_{[\rho }^{\;\;[\alpha }\delta _{\lambda
]}^{\beta ]}\right) \partial _{[\alpha }\eta _{\beta ]}\right.   \nonumber \\
&&\left. +\left( \partial _{[\rho }\hat{M}_{\lambda ]}+\hat{M}_{[\mu \nu
]}\right) \xi +\left( -\hat{M}_{[\rho }\delta _{\lambda ]}^{\alpha
}+\partial _{[\rho }\tilde{M}_{\lambda ]}^{\;\;\alpha }-4\bar{M}_{[\rho
\lambda ]}^{\;\;\alpha }\right) \partial _{\alpha }\xi \right]   \nonumber \\
&&+\bar{C}\left[ \varepsilon ^{\mu \nu \rho \lambda }\gamma _{5}\gamma _{\mu
}\partial _{\nu }\left( -\tilde{M}_{\rho \alpha }\mathcal{K}_{\lambda
}^{\alpha }+\bar{M}_{\rho }^{\;\;\alpha \beta }\partial _{[\alpha }h_{\beta
]\lambda }+\tfrac{1}{2}\hat{M}_{\rho \alpha }h_{\lambda }^{\alpha }\right)
\right.   \nonumber \\
&&\left. +\tfrac{1}{2}\mathcal{W}^{\mu \alpha \nu \beta }\partial _{[\mu
}M_{\alpha ][\nu ,\beta ]}\right] .  \label{3.30}
\end{eqnarray}
If we look at (\ref{3.27}), we observe that the existence of $a_{0}^{(\text{%
int})}$ as solution to the equation $\delta a_{1}^{(\text{int})}+\gamma
a_{0}^{(\text{int})}=\partial _{\mu }\stackrel{(0)}{m}^{(\text{int})\mu }$
requires that
\begin{equation}
f_{0}=0.  \label{3.30a}
\end{equation}
This observation leads to the following identities that must be satisfied by
the various fermionic spinor-tensors
\begin{equation}
\partial _{[\rho }\hat{M}_{\lambda ]\alpha }=0,  \label{3.31}
\end{equation}
\begin{equation}
\partial _{[\rho }\bar{M}_{\lambda ]}^{\;\;\alpha \beta }+\tfrac{1}{4}\hat{M}%
_{[\rho }^{\;\;[\alpha }\delta _{\lambda ]}^{\beta ]}=0,  \label{3.32}
\end{equation}
\begin{equation}
\partial _{[\rho }\hat{M}_{\lambda ]}+\hat{M}_{[\mu \nu ]}=0,  \label{3.33}
\end{equation}
\begin{equation}
-\hat{M}_{[\rho }\delta _{\lambda ]}^{\alpha }+\partial _{[\rho }\tilde{M}%
_{\lambda ]}^{\;\;\alpha }-4\bar{M}_{[\rho \lambda ]}^{\;\;\alpha }=0,
\label{3.34}
\end{equation}
\begin{eqnarray}
&&\varepsilon ^{\mu \nu \rho \lambda }\gamma _{5}\gamma _{\mu }\partial
_{\nu }\left( -\tilde{M}_{\rho \alpha }\mathcal{K}_{\lambda }^{\alpha }+\bar{%
M}_{\rho }^{\;\;\alpha \beta }\partial _{[\alpha }h_{\beta ]\lambda }+\tfrac{%
1}{2}\hat{M}_{\rho \alpha }h_{\lambda }^{\alpha }\right)   \nonumber \\
&&+\tfrac{1}{2}\mathcal{W}^{\mu \alpha \nu \beta }\partial _{[\mu }M_{\alpha
][\nu ,\beta ]}=0.  \label{3.34a}
\end{eqnarray}
The solutions to the equations (\ref{3.31})--(\ref{3.34}) can be expressed
in terms of some arbitrary, fermionic, spinor and gauge-invariant objects,
like
\begin{eqnarray}
\hat{M}_{\mu \alpha } &=&\partial _{\mu }\hat{N}_{\alpha },  \label{3.35a} \\
\hat{M}_{\alpha } &=&-\hat{N}_{\alpha }+\partial _{\alpha }\hat{M},
\label{3.35b} \\
\bar{M}_{\mu }^{\;\;\alpha \beta } &=&-\tfrac{1}{4}\hat{N}^{[\alpha }\delta
_{\mu }^{\beta ]}+\partial _{\mu }\bar{M}^{\alpha \beta },  \label{3.35c} \\
\tilde{M}_{\mu \nu } &=&4\bar{M}_{\mu \nu }+\hat{M}\sigma _{\mu \nu
}+\partial _{\mu }\tilde{M}_{\alpha },  \label{3.35d}
\end{eqnarray}
where $\bar{M}_{\mu \nu }$ is antisymmetric. Using the solutions (\ref{3.35a}%
)--(\ref{3.35d}) in (\ref{3.34a}), we then find
\begin{eqnarray}
&&\tfrac{1}{2}\mathcal{W}^{\mu \nu \alpha \beta }\left[ \partial _{[\mu
}M_{\nu ][\alpha ,\beta ]}+\gamma _{5}\gamma ^{\rho }\partial ^{\lambda
}\left( \varepsilon _{\mu \nu \rho \lambda }\bar{M}_{\alpha \beta
}+\varepsilon _{\alpha \beta \rho \lambda }\bar{M}_{\mu \nu }\right) \right]
\nonumber \\
&&+\tfrac{1}{2}\varepsilon ^{\mu \nu \rho \lambda }\gamma _{5}\gamma _{\mu
}\left( 4\bar{M}_{\nu \alpha }+\partial _{\nu }\tilde{M}_{\alpha }\right)
\partial _{[\rho }\mathcal{K}_{\lambda ]}^{\;\;\alpha }=0.  \label{3.35e}
\end{eqnarray}
This last equality yields
\begin{equation}
\partial _{[\mu }M_{\nu ][\alpha ,\beta ]}+\gamma _{5}\gamma ^{\rho
}\partial ^{\lambda }\left( \varepsilon _{\mu \nu \rho \lambda }\bar{M}%
_{\alpha \beta }+\varepsilon _{\alpha \beta \rho \lambda }\bar{M}_{\mu \nu
}\right) =0,  \label{3.35f}
\end{equation}
\begin{equation}
4\bar{M}_{\nu \alpha }+\partial _{\nu }\tilde{M}_{\alpha }=0.  \label{3.35g}
\end{equation}
If we take the symmetric part of (\ref{3.35g}) and perform some simple
computation, we get
\begin{equation}
\tilde{M}_{\alpha }=0,  \label{3.35h}
\end{equation}
which then produces
\begin{equation}
\bar{M}_{\nu \alpha }=0.  \label{3.35ha}
\end{equation}
Replacing the results (\ref{3.35h})--(\ref{3.35ha}) into (\ref{3.35f}), we
obtain the equation
\begin{equation}
\partial _{[\mu }M_{\nu ][\alpha ,\beta ]}=0.  \label{3.35i}
\end{equation}
Putting together the relations (\ref{3.35a})--(\ref{3.35d}) and (\ref{3.35h}%
)--(\ref{3.35i}), we conclude that the solutions to the equations (\ref{3.31}%
)--(\ref{3.34a}) can be written like
\begin{eqnarray}
\hat{M}_{\mu \alpha } &=&\partial _{\mu }\hat{N}_{\alpha },  \label{3.36a} \\
\hat{M}_{\alpha } &=&-\hat{N}_{\alpha }+\partial _{\alpha }\hat{M},
\label{3.36b} \\
\bar{M}_{\mu }^{\;\;\alpha \beta } &=&-\tfrac{1}{4}\hat{N}^{[\alpha }\delta
_{\mu }^{\beta ]},  \label{3.36c} \\
\tilde{M}_{\mu \nu } &=&\hat{M}\sigma _{\mu \nu },  \label{3.36d} \\
M_{\mu \nu } &=&\partial _{(\mu }M_{\nu )},  \label{3.36e}
\end{eqnarray}
in terms of some arbitrary, fermionic, spinor, gauge-invariant objects $\hat{%
N}_{\alpha }$, $\hat{M}$ and $M_{\nu }$. Inserting the solutions (\ref{3.36a}%
)--(\ref{3.36e}) into (\ref{3.26}), after some computation we arrive at
\begin{equation}
a_{1}^{(\text{int})}=\partial _{\mu }n^{\mu }+sq+\gamma r,  \label{3.36}
\end{equation}
where the following notations were employed
\begin{eqnarray}
n^{\mu } &=&\psi ^{*\mu }\left( \hat{M}\xi +\hat{N}^{\nu }\eta _{\nu
}\right) +2h^{*\mu \nu }\bar{C}M_{\nu },  \label{3.37a} \\
r &=&-\tfrac{1}{2}\psi ^{*\mu }\hat{N}^{\nu }h_{\mu \nu }+2h^{*\mu \nu }\bar{%
\psi}_{\mu }M_{\nu },  \label{3.37b} \\
q &=&-C^{*}\left( \hat{M}\xi +\hat{N}^{\nu }\eta _{\nu }\right) -2\eta
^{*\mu }\bar{C}M_{\mu }.  \label{3.37c}
\end{eqnarray}
As we have previously discussed, the solution to $a_{1}^{(\text{int})}$ is
unique on the one hand up to $\gamma $-trivial contributions (since it
represents the component of highest antighost number from the first-order
deformation, as it can be seen from the expansion (\ref{3.25})) and, on the
other hand, up to $s$-exact modulo $d$ terms (since it belongs to the
first-order deformation), so we can choose
\[
a_{1}^{(\text{int})}=0.
\]
In conclusion, there are no nontrivial, cross-coupling first-order
deformations that end at antighost number one. It is important to note that
the absence of consistent, nontrivial first-order deformations that end at
antighost number one also emerged \emph{without} imposing any restriction on
the derivative order of the interacting Lagrangian.

\subsection{Case $I=0$\label{i0}}

At this stage, we are left with a sole possibility, namely, that $a^{(\text{%
int})}$ reduces to its antighost number zero component ($I=0$)
\begin{equation}
a^{(\text{int})}=a_{0}^{(\text{int})},  \label{3.38}
\end{equation}
which is subject to the equation
\begin{equation}
\gamma a_{0}^{(\text{int})}=\partial _{\mu }\stackrel{(0)}{m}^{(\text{int}%
)\mu },  \label{3.39}
\end{equation}
for some local $\stackrel{(0)}{m}^{(\text{int})\mu }$. There are two main
types of solutions to (\ref{3.39}). The first type, to be denoted by $%
a_{0}^{\prime (\text{int})}$, corresponds to $\stackrel{(0)}{m}^{(\text{int}%
)\mu }=0$ and is given by gauge-invariant, nonintegrated densities
constructed from the original fields and their spacetime derivatives, which,
according to (\ref{3.10}), are given by
\begin{equation}
a_{0}^{\prime (\text{int})}=a_{0}^{\prime (\text{int})}\left( \left[
\partial _{[\mu }\psi _{\nu ]}\right] ,\left[ \mathcal{W}_{\mu \nu \alpha
\beta }\right] \right) ,  \label{3.41}
\end{equation}
up to the condition that they effectively describe cross-couplings between
the two types of fields and cannot be written in a divergence-like form. At
this point we invoke the hypothesis on the preservation of the number of
derivatives on each field, which means here that the following two
requirements are simultaneously satisfied: (i) the derivative order of the
equations of motion on each field is the same for the free and respectively
for the interacting theory; (ii) the maximum number of derivatives in the
interaction vertices is equal to four, i.e. the maximum number of
derivatives from the free lagrangian. This further yields the trivial
solution
\begin{equation}
a_{0}^{\prime (\text{int})}=0.  \label{3.41a}
\end{equation}
If we however relax the derivative-order condition, we can find nonvanishing
solutions of the type (\ref{3.41})\footnote{\label{note1}An example of a
possible solution of the form (\ref{3.41}) is represented by the cubic
vertex $a_{0}^{(\text{int})}=\mathcal{W}^{\mu \nu \alpha \beta }\left(
\partial _{[\mu }\bar{\psi}_{\nu ]}\right) \gamma _{5}\partial _{[\alpha
}\psi _{\beta ]}$.}. In conclusion, the condition on the number of
derivatives prevents the appearance of the solutions of the first type.

The second kind of solutions, to be denoted by $a_{0}^{\prime \prime (\text{%
int})}$, is associated with $\stackrel{(0)}{m}^{(\text{int})\mu }\neq 0$ in (%
\ref{3.39})
\begin{equation}
\gamma a_{0}^{\prime \prime (\text{int})}=\partial _{\mu }\stackrel{(0)}{m}%
^{(\text{int})\mu }.  \label{3.41c}
\end{equation}
In order to solve the equation (\ref{3.41c}) we recall the requirement that $%
a_{0}^{\prime \prime (\text{int})}$ may contain at most four derivatives of
the fields. Thus, it is useful to decompose $a_{0}^{\prime \prime (\text{int}%
)}$ according to the number of derivatives into
\begin{equation}
a_{0}^{\prime \prime (\text{int})}=\lambda _{0}+\lambda _{1}+\lambda
_{2}+\lambda _{3}+\lambda _{4},  \label{3.41b}
\end{equation}
where $\left( \lambda _{i}\right) _{i=\overline{0,4}}$ contains $i$
derivatives. Substituting (\ref{3.41b}) in (\ref{3.41c}), it follows that
the latter equation becomes equivalent to two independent equations, one for
each derivative order
\begin{eqnarray}
\gamma \lambda _{0} &=&\partial _{\mu }p^{\mu },  \label{3.41d} \\
\gamma \lambda _{1} &=&\partial _{\mu }q^{\mu },  \label{3.41e} \\
\gamma \lambda _{k} &=&\partial _{\mu }l_{k}^{\mu },\;k=2,3,4.
\label{3.41ex}
\end{eqnarray}
As $\lambda _{0}$ has no derivatives, we find that
\begin{equation}
\gamma \lambda _{0}=\frac{\partial ^{R}\lambda _{0}}{\partial \psi _{\mu }}%
\partial _{\mu }C+\frac{\partial \lambda _{0}}{\partial h_{\mu \nu }}\left(
\partial _{(\mu }\eta _{\nu )}+2\sigma _{\mu \nu }\xi \right) .
\label{3.41f}
\end{equation}
The right-hand side of (\ref{3.41f}) can be written like in the right-hand
side of (\ref{3.41d}) if
\begin{equation}
\partial _{\mu }\left( \frac{\partial ^{R}\lambda _{0}}{\partial \psi _{\mu }%
}\right) =0,\qquad \partial _{\mu }\left( \frac{\partial \lambda _{0}}{%
\partial h_{\mu \nu }}\right) =0,\qquad \frac{\partial \lambda _{0}}{%
\partial h_{\mu \nu }}\sigma _{\mu \nu }=0.  \label{3.41g}
\end{equation}
It is easy to see that the only solution to the all of the above equations
is a (real) constant, which can always be taken to vanish. Let us analyze
now the equation (\ref{3.41e}). We denote the Euler-Lagrange derivatives of $%
\lambda _{1}$ by
\begin{equation}
\bar{A}^{\alpha \beta }=\frac{\delta \lambda _{1}}{\delta h_{\alpha \beta }}%
,\qquad \bar{B}^{\mu }=\frac{\delta ^{R}\lambda _{1}}{\delta \psi _{\mu }},
\label{3.44}
\end{equation}
and ask that (the symmetric tensor) $\bar{A}^{\alpha \beta }$ contains at
least two massless Rarita-Schwinger fields and that $\bar{B}^{\mu }$
includes at least one Weyl graviton (in order to enforce the existence of
effective cross-couplings). At the same time, $\bar{A}^{\alpha \beta }$ and $%
\bar{B}^{\mu }$ are precisely of order one in the field derivatives, with
both $\bar{A}^{\alpha \beta }$ and $\bar{B}^{\mu }$ real objects (the first
is bosonic and the second fermionic), and, moreover, $\bar{B}^{\mu }$ is a
spinor-vector. Using the definitions (\ref{16}), it follows that the
equation (\ref{3.41e}) further restricts $\bar{A}^{\alpha \beta }$ and $\bar{%
B}^{\mu }$ to satisfy the equations
\begin{equation}
\partial _{\mu }\bar{B}^{\mu }=0,  \label{3.45a}
\end{equation}
\begin{equation}
\partial _{\alpha }\bar{A}^{\alpha \beta }=0,\qquad \bar{A}^{\alpha \beta
}\sigma _{\alpha \beta }=0.  \label{3.45b}
\end{equation}
The solutions to the above equations are
\begin{eqnarray}
\bar{B}^{\mu } &=&\partial _{\nu }\bar{B}^{\nu \mu },\qquad \bar{B}^{\mu \nu
}=-\bar{B}^{\nu \mu },  \label{3.45z} \\
\bar{A}^{\alpha \beta } &=&\partial _{\rho }\bar{A}^{\rho \alpha \beta
},\qquad \bar{A}^{\rho \alpha \beta }=-\bar{A}^{\alpha \rho \beta },\qquad
\bar{A}^{\rho \alpha \beta }\sigma _{\alpha \beta }=0,  \label{3.45z1}
\end{eqnarray}
where the antisymmetric tensors $\bar{B}^{\mu \nu }$ and $\bar{A}^{\rho
\alpha \beta }$ depend only on the original undifferentiated fields. The
tensors $\bar{B}^{\mu \nu }$ and $\bar{A}^{\rho \alpha \beta }$ have the
same properties (Grassmann parity, reality, spinor-like behavior) like $%
\bar{B}^{\mu }$ and respectively $\bar{A}^{\alpha \beta }$. We insist on the
fact that a solution of the type $\bar{A}^{\mu \nu }=\partial _{\alpha
}\partial _{\beta }D^{\mu \alpha \nu \beta }$, with $D^{\mu \alpha \nu \beta
}$ possessing the symmetry properties of the Riemann tensor, is not allowed
in our case due to the hypothesis on the derivative order, and hence (\ref
{3.45z1}) is the most general solution to the equations (\ref{3.45b}) in
this case.

Let $N$ be a derivation in the algebra of the fields and of their spacetime
derivatives, that counts the number of the fields and of their derivatives,
defined by
\begin{eqnarray}
N &=&\sum\limits_{k\geq 0}\left[ \frac{\partial ^{R}}{\partial \left(
\partial _{\mu _{1}}\cdots \partial _{\mu _{k}}\psi _{\alpha }\right) }%
\left( \partial _{\mu _{1}}\cdots \partial _{\mu _{k}}\psi _{\alpha }\right)
\right.  \nonumber \\
&&\left. +\left( \partial _{\mu _{1}}\cdots \partial _{\mu _{k}}h_{\alpha
\beta }\right) \frac{\partial }{\partial \left( \partial _{\mu _{1}}\cdots
\partial _{\mu _{k}}h_{\alpha \beta }\right) }\right] .  \label{opn}
\end{eqnarray}
Then, it is easy to see that for every nonintegrated density $u$ we have
that
\begin{equation}
Nu=\frac{\delta ^{R}u}{\delta \psi _{\alpha }}\psi _{\alpha }+\frac{\delta u%
}{\delta h_{\alpha \beta }}h_{\alpha \beta }+\partial _{\mu }s^{\mu },
\label{3.47a}
\end{equation}
where $\delta ^{R}u/\delta \psi _{\alpha }$ and $\delta u/\delta h_{\alpha
\beta }$ denote the Euler-Lagrange derivatives of $u$ with respect to $\psi
_{\alpha }$ and respectively to $h_{\alpha \beta }$. If $u^{(l)}$ is a
homogeneous polynomial of order $l>0$ in the fields and their spacetime
derivatives, then $Nu^{(l)}=lu^{(l)}$. Using (\ref{3.44}), (\ref{3.45z})--(%
\ref{3.45z1}) and (\ref{3.47a}) we infer that
\begin{equation}
N\lambda _{1}=-\tfrac{1}{2}\left( \bar{B}^{\mu \nu }\partial _{[\mu }\psi
_{\nu ]}+\bar{A}^{\rho \alpha \beta }\partial _{[\rho }h_{\alpha ]\beta
}\right) +\partial _{\mu }\bar{s}^{\mu }.  \label{3.48}
\end{equation}
Now, we expand $\lambda _{1}$ like
\begin{equation}
\lambda _{1}=\sum\limits_{l>0}\lambda _{1}^{(l)},  \label{3.48a}
\end{equation}
where $N\lambda _{1}^{(l)}=l\lambda _{1}^{(l)}$, such that
\begin{equation}
N\lambda _{1}=\sum\limits_{l>0}l\lambda _{1}^{(l)}.  \label{3.47b}
\end{equation}
Comparing the relation (\ref{3.48}) with (\ref{3.47b}), we conclude that $%
\bar{B}^{\mu \nu }$ and $\bar{A}^{\rho \alpha \beta }$ inherit some
decompositions similar to (\ref{3.48a})
\begin{equation}
\bar{B}^{\mu \nu }=\sum\limits_{l>0}\bar{B}_{(l-1)}^{\mu \nu },\qquad \bar{A}%
^{\rho \alpha \beta }=\sum\limits_{l>0}\bar{A}_{(l-1)}^{\rho \alpha \beta }.
\label{wx}
\end{equation}
Inserting (\ref{wx}) in (\ref{3.48}) and comparing the resulting expression
with (\ref{3.47b}) we deduce that
\begin{equation}
\lambda _{1}^{(l)}=-\tfrac{1}{2l}\left( \bar{B}_{(l-1)}^{\mu \nu }\partial
_{[\mu }\psi _{\nu ]}+\bar{A}_{(l-1)}^{\rho \alpha \beta }\partial _{[\rho
}h_{\alpha ]\beta }\right) +\partial _{\mu }\bar{s}_{(l)}^{\mu }.
\label{wx1}
\end{equation}
Replacing (\ref{wx1}) in (\ref{3.48a}), we find that
\begin{equation}
\lambda _{1}=-\tfrac{1}{2}\left( B^{\mu \nu }\partial _{[\mu }\psi _{\nu
]}+A^{\rho \alpha \beta }\partial _{[\rho }h_{\alpha ]\beta }\right)
+\partial _{\mu }z^{\mu },  \label{wx2}
\end{equation}
where
\begin{equation}
B^{\mu \nu }=\sum\limits_{l>0}\tfrac{1}{l}\bar{B}_{(l-1)}^{\mu \nu },\qquad
A^{\rho \alpha \beta }=\sum\limits_{l>0}\tfrac{1}{l}\bar{A}_{(l-1)}^{\rho
\alpha \beta }.  \label{wx3}
\end{equation}
Using (\ref{wx2}) we obtain that
\begin{eqnarray}
\gamma \lambda _{1} &=&-\tfrac{1}{2}\left\{ \bar{C}\partial _{\lambda
}\left( \frac{\partial ^{L}B^{\mu \nu }}{\partial \bar{\psi}_{\lambda }}%
\partial _{[\mu }\psi _{\nu ]}-\frac{\partial ^{L}A^{\mu \nu \rho }}{%
\partial \bar{\psi}_{\lambda }}\partial _{[\mu }h_{\nu ]\rho }\right) \right.
\nonumber \\
&&+2\left( \frac{\partial B^{\mu \nu }}{\partial h_{\alpha \beta }}\partial
_{[\mu }\psi _{\nu ]}+\frac{\partial A^{\mu \nu \rho }}{\partial h_{\alpha
\beta }}\partial _{[\mu }h_{\nu ]\rho }\right) \sigma _{\alpha \beta }\xi
\nonumber \\
&&-2\left[ \partial _{\alpha }\left( \frac{\partial B^{\mu \nu }}{\partial
h_{\alpha \beta }}\partial _{[\mu }\psi _{\nu ]}+\frac{\partial A^{\mu \nu
\rho }}{\partial h_{\alpha \beta }}\partial _{[\mu }h_{\nu ]\rho }\right)
\right] \eta _{\beta }  \nonumber \\
&&\left. -\left( \partial _{\beta }A^{\rho \alpha \beta }\right) \partial
_{[\rho }\eta _{\alpha ]}\right\} +\partial _{\mu }t^{\mu }.  \label{3.49}
\end{eqnarray}
Comparing (\ref{3.49}) with (\ref{3.41e}) and taking into account the fact
that $\bar{C}$, $\xi $, $\eta _{\beta }$ and $\partial _{[\rho }\eta
_{\alpha ]}$ are independent elements of pure ghost number equal to one of
the basis in the space of polynomials in the ghosts, we find that the
tensors $B^{\mu \nu }$ and $A^{\rho \alpha \beta }$ are restricted to
fulfill the conditions:
\begin{eqnarray}
\partial _{\lambda }\left( \frac{\partial ^{L}B^{\mu \nu }}{\partial \bar{%
\psi}_{\lambda }}\partial _{[\mu }\psi _{\nu ]}-\frac{\partial ^{L}A^{\mu
\nu \rho }}{\partial \bar{\psi}_{\lambda }}\partial _{[\mu }h_{\nu ]\rho
}\right) &=&0,  \label{3.50a} \\
\left( \frac{\partial B^{\mu \nu }}{\partial h_{\alpha \beta }}\partial
_{[\mu }\psi _{\nu ]}+\frac{\partial A^{\mu \nu \rho }}{\partial h_{\alpha
\beta }}\partial _{[\mu }h_{\nu ]\rho }\right) \sigma _{\alpha \beta } &=&0,
\label{3.50b} \\
\partial _{\alpha }\left( \frac{\partial B^{\mu \nu }}{\partial h_{\alpha
\beta }}\partial _{[\mu }\psi _{\nu ]}+\frac{\partial A^{\mu \nu \rho }}{%
\partial h_{\alpha \beta }}\partial _{[\mu }h_{\nu ]\rho }\right) &=&0,
\label{3.50c} \\
\partial _{\beta }A^{\rho \alpha \beta } &=&0.  \label{3.50d}
\end{eqnarray}
Since $A^{\rho \alpha \beta }$ are nonderivative functions, from the last
condition we deduce that they are constant. By covariance arguments they
must vanish
\begin{equation}
A^{\rho \alpha \beta }=0.  \label{3.50e}
\end{equation}
Inserting (\ref{3.50e}) in (\ref{3.50a})--(\ref{3.50c}) we arrive at
\begin{eqnarray}
\partial _{\lambda }\left( \frac{\partial ^{L}B^{\mu \nu }}{\partial \bar{%
\psi}_{\lambda }}\partial _{[\mu }\psi _{\nu ]}\right) &=&0,  \label{3.51a}
\\
\sigma _{\alpha \beta }\frac{\partial B^{\mu \nu }}{\partial h_{\alpha \beta
}}\partial _{[\mu }\psi _{\nu ]} &=&0,  \label{3.51} \\
\partial _{\alpha }\left( \frac{\partial B^{\mu \nu }}{\partial h_{\alpha
\beta }}\partial _{[\mu }\psi _{\nu ]}\right) &=&0.  \label{3.51m}
\end{eqnarray}
The equation (\ref{3.51a}) implies that
\begin{equation}
\frac{\partial ^{L}B^{\mu \nu }}{\partial \bar{\psi}_{\lambda }}\partial
_{[\mu }\psi _{\nu ]}=\partial _{\mu }S^{\mu \lambda },\qquad S^{\mu \lambda
}=-S^{\lambda \mu },  \label{3.52a}
\end{equation}
for some spinor-tensor $S^{\mu \lambda }$. By direct computation we find
\begin{eqnarray}
\frac{\partial ^{L}B^{\mu \nu }}{\partial \bar{\psi}_{\lambda }}\partial
_{[\mu }\psi _{\nu ]} &=&\partial _{\mu }\left( \frac{\partial ^{L}B^{\mu
\nu }}{\partial \bar{\psi}_{\lambda }}\psi _{\nu }-\frac{\partial
^{L}B^{\lambda \nu }}{\partial \bar{\psi}_{\mu }}\psi _{\nu }\right)
\nonumber \\
&&+\partial _{\mu }\left( \frac{\partial ^{L}B^{\mu \nu }}{\partial \bar{\psi%
}_{\lambda }}\psi _{\nu }+\frac{\partial ^{L}B^{\lambda \nu }}{\partial \bar{%
\psi}_{\mu }}\psi _{\nu }\right)  \nonumber \\
&&-2\left( \partial _{\mu }\frac{\partial ^{L}B^{\mu \nu }}{\partial \bar{%
\psi}_{\lambda }}\right) \psi _{\nu }.  \label{3.52b}
\end{eqnarray}
Comparing (\ref{3.52a}) with (\ref{3.52b}) we infer that the last two terms
in the right-hand side of (\ref{3.52b}) must vanish
\begin{eqnarray}
\partial _{\mu }\left( \frac{\partial ^{L}B^{\mu \nu }}{\partial \bar{\psi}%
_{\lambda }}\right) &=&0,  \label{3.52c} \\
\partial _{\mu }\left( \frac{\partial ^{L}B^{\mu \nu }}{\partial \bar{\psi}%
_{\lambda }}\psi _{\nu }+\frac{\partial ^{L}B^{\lambda \nu }}{\partial \bar{%
\psi}_{\mu }}\psi _{\nu }\right) &=&0.  \label{3.52d}
\end{eqnarray}
As $B^{\mu \nu }$ contains no derivatives, the equation (\ref{3.52c}) gives
\begin{equation}
\frac{\partial ^{L}B^{\mu \nu }}{\partial \bar{\psi}_{\lambda }}=C^{\lambda
\mu \nu },\qquad C^{\lambda \mu \nu }=-C^{\lambda \nu \mu },  \label{3.52e}
\end{equation}
for some constant matrices $C^{\lambda \mu \nu }$. Substituting (\ref{3.52e}%
) into (\ref{3.52d}), we obtain the equation $\partial _{\mu }\left( \left(
C^{\lambda \mu \nu }+C^{\mu \lambda \nu }\right) \psi _{\nu }\right) =0$,
which further implies
\begin{equation}
C^{\lambda \mu \nu }=-C^{\mu \lambda \nu },  \label{3.52f}
\end{equation}
so the objects $C^{\lambda \mu \nu }$ are completely antisymmetric in their
Lorentz indices. In consequence, from (\ref{3.52e}) we get
\begin{equation}
B^{\mu \nu }=\bar{\psi}_{\lambda }C^{\lambda \mu \nu }.  \label{3.52g}
\end{equation}
From (\ref{3.52g}) we find that $\partial B^{\mu \nu }/\partial h_{\alpha
\beta }=0$, so the solution (\ref{3.52g}) verifies also the equations (\ref
{3.51}) and (\ref{3.51m}). In the meantime, the general form of the constant
matrices $C^{\lambda \mu \nu }$ reads as $C^{\lambda \mu \nu }=k\gamma
^{[\lambda }\gamma ^{\mu }\gamma ^{\nu ]}$, with $k$ an arbitrary numerical
constant, such that
\begin{equation}
B^{\mu \nu }=k\bar{\psi}_{\lambda }\gamma ^{[\lambda }\gamma ^{\mu }\gamma
^{\nu ]}.  \label{3.52h}
\end{equation}
Introducing the solutions (\ref{3.50e}) and (\ref{3.52h}) in (\ref{wx2}), it
follows that
\begin{equation}
\lambda _{1}\propto \bar{\psi}_{\lambda }\gamma ^{[\lambda }\gamma ^{\mu
}\gamma ^{\nu ]}\partial _{[\mu }\psi _{\nu ]}.  \label{3.52}
\end{equation}

We are now left with investigating the solutions to the equations (\ref
{3.41ex}). Taking into account the hypothesis on the preservation of the
number of derivatives on each field, we obtain that
\begin{eqnarray}
\lambda _{2} &=&\lambda _{2}\left( \psi ,h,\partial h,\partial h\right) ,
\label{a06} \\
\lambda _{3} &=&\lambda _{3}\left( \psi ,h,\partial h,\partial h,\partial
h\right) ,  \label{a07} \\
\lambda _{4} &=&\lambda _{4}\left( \psi ,h,\partial h,\partial h,\partial
h,\partial h\right) ,  \label{a08}
\end{eqnarray}
which signifies that for the values $k=2,3,4$ the functions $\lambda _{k}$
depends only on the undifferentiated Rarita-Schwinger field, on the
undifferentiated Weyl field, and also on $k$ spacetime derivatives of order
one of the Weyl field. If we make the notations
\begin{equation}
\tilde{A}_{k}^{\alpha \beta }=\frac{\delta \lambda _{k}}{\delta h_{\alpha
\beta }},\qquad \tilde{B}_{k}^{\mu }=\frac{\partial ^{R}\lambda _{k}}{%
\partial \psi _{\mu }},\qquad k=2,3,4,  \label{a09}
\end{equation}
on behalf of the formulas (\ref{a06})--(\ref{a08}) we deduce that
\begin{eqnarray}
\tilde{B}_{2}^{\mu } &=&\tilde{B}_{2}^{\mu }\left( \psi ,h,\partial
h,\partial h\right) ,  \label{a010} \\
\tilde{B}_{3}^{\mu } &=&\tilde{B}_{3}^{\mu }\left( \psi ,h,\partial
h,\partial h,\partial h\right) ,  \label{a011} \\
\tilde{B}_{4}^{\mu } &=&\tilde{B}_{4}^{\mu }\left( \psi ,h,\partial
h,\partial h,\partial h,\partial h\right) .  \label{a012}
\end{eqnarray}
By means of the definitions (\ref{16}), we infer that the equations (\ref
{3.41ex}) are satisfied if the Euler-Lagrange derivatives $\tilde{A}%
_{k}^{\alpha \beta }$ and $\tilde{B}_{k}^{\mu }$\ are subject to the
equations
\begin{equation}
\partial _{\mu }\tilde{B}_{k}^{\mu }=0,  \label{a013}
\end{equation}
\begin{equation}
\partial _{\alpha }\tilde{A}_{k}^{\alpha \beta }=0,\qquad \tilde{A}%
_{k}^{\alpha \beta }\sigma _{\alpha \beta }=0.  \label{a014}
\end{equation}
The solutions to the equations (\ref{a013}) are given by
\begin{equation}
\tilde{B}_{k}^{\mu }=\partial _{\nu }\tilde{B}_{k}^{\nu \mu },\qquad \tilde{B%
}_{k}^{\nu \mu }=-\tilde{B}_{k}^{\mu \nu },  \label{a016}
\end{equation}
while the solutions to (\ref{a014}) are not important in what follows and
will not be considered here. On the one hand, the relations (\ref{a010})--(%
\ref{a012}) show that $\tilde{B}_{k}^{\mu }$ do not depend on the
derivatives of the Rarita-Schwinger field, and, on the other hand, the
equations (\ref{a016}) imply that $\tilde{B}_{k}^{\mu }$ do depend on these
derivatives. As a consequence, the solutions to the equations (\ref{a013})
are merely constant, i.e.,
\begin{equation}
\tilde{B}_{k}^{\mu }=C_{k}^{\mu }.  \label{a017}
\end{equation}
Due to the fact that $\tilde{B}_{k}^{\mu }$ are spinor-vectors and since the
present field spectrum does not allow the construction of constant
spinor-vectors, it results that the sole solution to the equations (\ref
{a013}) is vanishing
\begin{equation}
\tilde{B}_{k}^{\mu }=0.  \label{a018}
\end{equation}
Replacing (\ref{a018}) in the latter relation in (\ref{a09}), we reach the
conclusion that the quantities $\left( \lambda _{k}\right) _{k=\overline{2,4}%
}$ do not depend on the Rarita-Schwinger field and, consequently, they
cannot describe cross-couplings between the Weyl graviton and the
Rarita-Schwinger field, as required, so we can take
\[
\lambda _{k}=0,\qquad k=2,3,4.
\]
In this manner we arrive at the final solution
\begin{equation}
a_{0}^{(\text{int})}=a_{0}^{\prime (\text{int})}+a_{0}^{\prime \prime (\text{%
int})}=\lambda _{1}\propto \bar{\psi}_{\lambda }\gamma ^{[\lambda }\gamma
^{\mu }\gamma ^{\nu ]}\partial _{[\mu }\psi _{\nu ]}  \label{a019}
\end{equation}
to the equation (\ref{3.39}).The above $a_{0}^{(\text{int})}$ does not
describe cross-couplings between the Weyl graviton and the Rarita-Schwinger
field. Moreover, it neither produces self-interactions of the
Rarita-Schwinger field since it is proportional with the free Lagrangian of
this theory, and, accordingly, it must be discarded from the first-order
deformation. Thus, there is no nontrivial possibility to couple the Weyl
graviton to the massless Rarita-Schwinger field by means of a first-order
deformation that reduces to its antighost number zero component under the
working hypotheses invoked in this paper. Thus, the conclusion of this
section is that the first-order deformation vanishes
\begin{equation}
S_{1}=0,  \label{a021}
\end{equation}
so the solutions to the higher-order deformation equations, (\ref{2.6})--(%
\ref{2.7}), etc., also vanish
\begin{equation}
S_{2}=S_{3}=\cdots =0.  \label{a020}
\end{equation}

\section{Conclusion}

In this paper we have investigated the cross-couplings that can be
introduced between the Weyl graviton and the massless Rarita-Schwinger field
from the BRST formalism point of view. Thus, under the general conditions of
locality, smoothness, (background) Lorentz invariance, Poincar\'{e}
invariance and preservation of the number of derivatives with respect to
each field (the last hypothesis was made only in antighost number zero), we
have proved that there are no such cross-couplings. The only deformations
that can be introduced in relation with the free model under study are
represented by the self-interactions of Weyl gravity, since there are no
consistent, nontrivial self-interactions of the massless Rarita-Schwinger
field (see, for instance, the consistency arguments invoked in~\cite{boulcqg}%
).

\section*{Acknowledgments}

Three of the authors (C.B., E.M.C., and S.C.S.) are partially supported by
the European Commission FP6 program MRTN-CT-2004-005104 and by the type A
grant 305/2004 with the Romanian National Council for Academic Scientific
Research (CNCSIS) and the Romanian Ministry of Education and Research (MEC).
One of the authors (A.C.L.) was supported by the World Federation of
Scientists (WFS).

\end{document}